\documentclass[conference,letterpaper]{IEEEtran}
\usepackage[T1]{fontenc}
\usepackage{newtxtext,newtxmath}
\usepackage{graphicx,amsmath,array,xcolor}
\usepackage{algorithm}
\usepackage[noend]{algpseudocode}
\usepackage[nocompress]{cite}
\usepackage[hidelinks]{hyperref}

\begin{document}
\title{An Optical KV Cache Network for\\
Disaggregated LLM Serving}
\author{\IEEEauthorblockN{Seongwon Yoon and Shimeng Yu}
\IEEEauthorblockA{School of Electrical and Computer Engineering, Georgia Institute of Technology, Atlanta, GA, USA\\
gabrielyoon@gatech.edu}}
\maketitle

\begin{abstract}
In disaggregated LLM serving, each request's KV cache moves from a prefill GPU to a decode GPU before the response
begins, and a coding-agent turn often moves the KV cache of its whole long context.
Behind an oversubscribed packet fabric, this transfer adds to the time to first token (TTFT).
An optical circuit switch (OCS) can give every transfer a full-rate circuit for less than an oversubscribed packet
fabric costs, but a mechanical OCS needs tens of milliseconds to reconfigure.
We hide this delay inside the prefill.
A reservation scheduler commits each circuit while the prefill is still computing, so the circuit is ready when the KV
cache is, and each GPU reaches the OCS through its own wavelength-multiplexed (WDM) optics on one fiber pair.
In simulations of GPU clusters on real serving traces with prefill times measured on H100 GPUs, this optical KV cache
network matches 400G packet fabrics with adaptive routing at 0.55 to 0.78 times the cost of a 2:1 fabric and tolerates
the 50\,ms reconfiguration delay of a 384-port mechanical OCS.
\end{abstract}

\section{Introduction}
Prefill/decode (P/D) disaggregation runs the two phases of LLM serving on separate GPU pools
\cite{distserve,splitwise,mooncake}.
Every request's KV cache then crosses the network and adds to TTFT unless the transfer ends within the prefill.
In our analysis of one P/D application in Alibaba's production trace~\cite{xmaas}, the hourly mean KV transfer time
is a median 1.10 times the prefill time over one week.
Deployed P/D systems choose the decode instance by load \cite{mooncake,dualpath,dynamo}, so an agent
turn~\cite{tracelab} that appends a tool result to a long context often moves the KV cache of the whole context, while
its prefill computes only the new tokens.
A non-blocking 400G fabric mostly hides this move, but operators oversubscribe the upper tiers
\cite{ark,hpn,metaroce}.
At half of 400G per GPU, the median migrated turn of Llama-3.1-70B on H100 GPUs adds 67\,ms to TTFT.

A non-blocking fabric costs 1.27 to 1.31 times a 2:1 fabric, and packet schedulers~\cite{ark,nuska} reduce contention
on the same links.
An OCS port draws far less power than a packet switch port and passes light without transceivers \cite{mixnet,opus},
so a circuit per transfer can give every GPU its full link rate for less than an oversubscribed fabric costs.
However, a mechanical OCS needs 10 to 50\,ms to reconfigure~\cite{mixnet,polatis6000n,polatis7000n}.
We hide this delay inside the prefill.
Earlier OCS fabrics reconfigure per training job, phase or workload
mode~\cite{mixnet,opus,mose,topoopt,tpuv4,silverstone}, while our circuits follow single KV transfers.
Our contributions are (1) an optical KV cache network with GPU-side WDM optics and the requirements it sets on them,
namely one fiber pair per 400G link, up to 50\,ms of reconfiguration and 2\,dB of link margin
(Section~\ref{sec:arch}, Table~\ref{tab:req}), (2) a reservation scheduler that commits the prefix circuit at
prefill start and the new-KV circuit at the last prefill chunk and reuses standing circuits (Section~\ref{sec:sched}),
and (3) simulations of H100 clusters on real serving traces with measured prefill times (Section~\ref{sec:eval}).

\begin{figure*}[t]
\centering
\includegraphics[width=\textwidth]{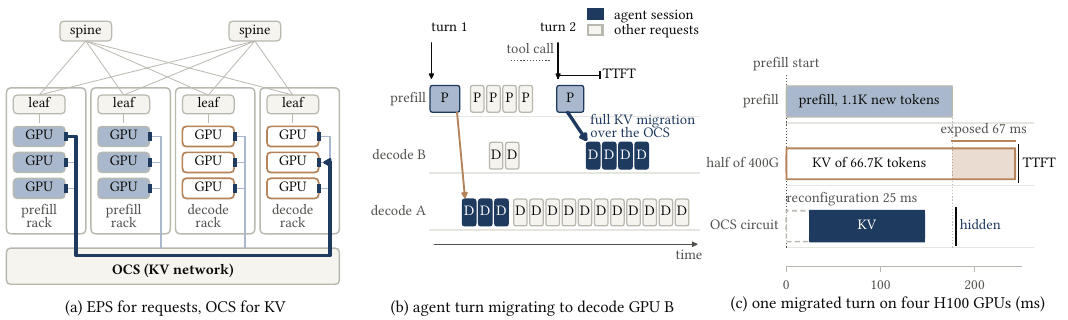}
\caption{Optical KV cache network. (a) Requests, output tokens and control stay on the host NICs and their leaf-spine
EPS, while the RDMA NIC of every GPU drives WDM optics into an OCS instead of a packet leaf. (b) After a tool call,
decode GPU A is busy with other requests, so turn 2 lands on decode GPU B and its full KV cache crosses the OCS on a
circuit committed during prefill. (c) Median migrated turn of Llama-3.1-70B on four H100 GPUs, modeled with the
measured prefill time. One circuit committed at prefill start reconfigures in 25\,ms and ends inside the 176\,ms
prefill.}
\label{fig:concept}
\end{figure*}

\section{Background}
\label{sec:background}
\subsection{Prefill, decode, and KV cache transfer}
Prefill reads the whole prompt at once, and decode generates one token at a time, both through the key and value
vectors that attention stores per token and layer, the KV cache.
For Llama-3.1-70B in BF16 it takes 320\,KiB per token~\cite{llama3}, so a 64K-token context occupies about 21\,GB.
Prefill is compute bound and decode is memory bound, so P/D disaggregation runs them on separate
pools~\cite{distserve,splitwise}, as in Mooncake, NVIDIA Dynamo and DeepSeek~\cite{mooncake,dualpath,dynamo}.
Mooncake streams the KV cache layer by layer during prefill, which hides the transfer while it ends within the
prefill.
The vLLM NIXL connector pulls it after prefill~\cite{vllmnixl}, which adds the whole transfer to TTFT, because decode
emits the leading token.

\subsection{Multi-turn agent sessions and KV cache migration}
Coding agents append a short input, such as a tool result, to a long shared context every turn.
In a trace of 4,300 coding-agent sessions from 43 developers~\cite{tracelab}, the steps that fit the 128K window of
Llama-3.1-70B have a median context of 66.7K tokens and a median new input of 1.1K tokens, against 88.8K and 1.6K for
main-agent turns in AgentX Claude Code sessions~\cite{agentx} and a 68K-token median prompt in GitHub
Copilot~\cite{copilot}.
With prefix caching, the prefill side keeps earlier turns' KV cache and computes only the new
tokens~\cite{vllm,sglang,preble}.
A turn migrates when the chosen decode instance does not hold the prefix, so its full KV cache crosses the network
before decode can respond.
Mooncake and DeepSeek send the full KV cache to a decode instance chosen by load every turn, and Dynamo ignores
prefix reuse when routing decode~\cite{mooncake,dualpath,dynamo}.
In llm-d runs on coding-agent traces, about half of the decode-side KV cache arrived from prefill~\cite{llmdglm}.
In Alibaba's production P/D telemetry~\cite{xmaas}, the least-utilized decode instance of a group changes in 42\% of
consecutive 5-minute windows.
Session affinity keeps a session on one decode instance, but 7.6\% of turns still miss decode memory in our modeled
replay, and a KV store still sends the KV cache over the fabric~\cite{dualpath}.

\subsection{Packet and circuit switching in the datacenter}
A GPU has three network paths.
NVLink joins the GPUs of a server, a 400G RDMA NIC per GPU joins GPUs across servers~\cite{hpn,metaroce}, and host
NICs carry requests, storage and management~\cite{hpn}.
Both NIC paths cross leaf and spine fat trees of electrical packet switches (EPS), whose upper tiers are often
oversubscribed~\cite{ark,hpn,metaroce}.
An EPS spreads flows by hashing (ECMP), which lets large transfers collide~\cite{hedera}, or by adaptive
routing~\cite{conga} under RDMA congestion control~\cite{dcqcn}.
An OCS steers light from an input fiber to an output fiber without converting it to an electrical
signal~\cite{helios,cthrough,lightwave}, as in Google's TPU and datacenter networks~\cite{tpuv4,jupiter,apollo}.
It does not inspect packets, so one port serves any line rate at 0.39\,W against 39.4\,W for a 400G packet switch
port~\cite{opus}, and at \$520 against \$583 to \$1,090~\cite{mixnet,opus}.
In exchange, a circuit joins one pair of ports at a time.
Mechanical switches, steered by MEMS mirrors or piezoelectric actuators, take up to 25\,ms to change it with 192 ports
and 50\,ms with 384 ports~\cite{mixnet,polatis6000n,polatis7000n}, and silicon photonic switches about a
microsecond~\cite{seok16}.

\section{Optical KV Cache Network}
\begin{figure}[t]
\centering
\includegraphics[width=\columnwidth]{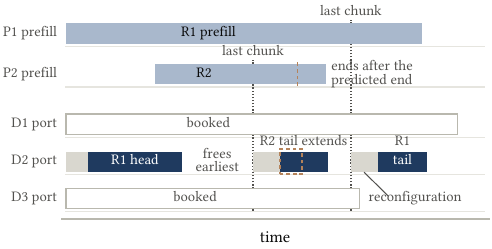}
\caption{Reservation calendar with last-chunk commit. R1 commits its head circuit at prefill start on the decode
port that frees earliest (D2) and its tail at the last chunk, so the reconfiguration delay (grey) overlaps that chunk.
The late tail of R2 extends into free port time.}
\label{fig:sched}
\end{figure}
\begin{algorithm}[t]
\caption{Reservation with last-chunk commit per request}
\label{alg:resv}
\footnotesize
\begin{algorithmic}[1]
\Statex \textbf{Admission}
\State $\hat T \gets$ prefill time predicted by the cost model
\State $B_h, B_t \gets$ KV bytes of the prefix and of the new tokens
\State split into head $B_h$ and tail $B_t$ only \textbf{if} $(B_h+B_t)/R+\tau \le 0.8\,\hat T$
\Statex \textbf{Prefill start} at $t_0$
\State $d \gets$ decode port free earliest, preferring a standing circuit
\State $\tau_h \gets 0$ \textbf{if} the circuit to $d$ stands \textbf{else} $\tau$
\State book head of length $\tau_h+B_h/R$ in the earliest gap after $t_0$ free on both ports
\Statex \textbf{Last prefill chunk} at $t_1$
\State $\hat E \gets$ prefill end re-predicted from observed progress
\State book tail on $d$ to end at $\hat E+T_\ell+0.1(\hat E-t_1)$
\Statex \textbf{Prefill end} at $E$
\If{the tail circuit is not up} move the tail to the earliest free gap \EndIf
\If{a burst outlasts its window} extend it into free port time, displacing bookings without ready data, else cut
it and re-book the rest \EndIf
\State keep the circuit up for reuse
\end{algorithmic}
\end{algorithm}
\begin{figure*}[t]
\centering
\includegraphics[width=\textwidth]{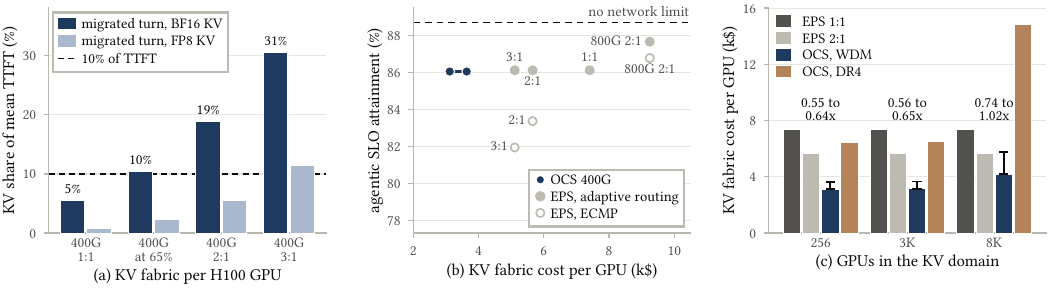}
\caption{(a) KV share of mean TTFT with each GPU at its share of 400G, Llama-3.1-70B on four H100 GPUs (modeled,
layer-wise), chat turns below 0.1\%. (b) Agentic SLO attainment against cost per GPU at 256 H100 GPUs (simulated,
pull), the OCS at one or two ports per fiber pair. (c) KV fabric cost per GPU at 400G with the prices
of~\cite{mixnet}. Whiskers reach two ports per fiber pair, ratios to EPS 2:1 above each group, three stages at 8K and
DR4 on eight ports.}
\label{fig:main}
\end{figure*}

\subsection{Architecture}
\label{sec:arch}
Requests, output tokens and control stay on the host NICs and their leaf-spine EPS, while the 400G RDMA NIC of each
GPU connects to an OCS instead of a packet leaf switch and carries the KV cache of every migrated turn
(Fig.~\ref{fig:concept}(a), (b)).
Each prefill and decode GPU drives one 400G link of pluggable or co-packaged WDM optics~\cite{ayarofc21} into the
OCS, 45\,GB/s at the 90\% efficiency we assume for every fabric.
A fiber OCS switches fibers, not wavelengths, so the GPU-side optics set its size.
A 400GBASE-FR4 link carries four wavelengths on one fiber pair, while a parallel-fiber 400GBASE-DR4 link occupies
four pairs~\cite{oifcpo} and needs four times the switch ports.
Eight 384-port planes, each joining one GPU position of every 8-GPU node, reach 3,072 GPUs in one
stage~\cite{polatis7000n}, so KV moves between equal GPU positions.

\subsection{Reservation scheduler}
\label{sec:sched}
The controller keeps one calendar per instance, listing the booked windows of its OCS ports and the peer each port
points to.
A window holds the reconfiguration delay, skipped when the circuit already stands toward that peer, and the
transfer.
A booking takes the earliest gap free on both the prefill and the decode port, so windows never overlap.
Algorithm~\ref{alg:resv} lists the four events of a request (Fig.~\ref{fig:sched}), with $R$ the circuit rate,
$\tau$ the reconfiguration delay and $T_\ell$ one prefill layer.
The head of each transfer carries the prefix KV, which the prefill GPU already holds and only reads, so it does not
depend on the tail, which carries the KV of the new tokens as their layers finish.
We commit the head at prefill start, when its start time is exact, and the tail when the last prefill chunk of up to
2,048 tokens starts, which leaves only the error of the remaining prefill.
A booking at admission would instead absorb the error of every prefill queued ahead.
Preferring a standing circuit (line 4) makes the decode choice circuit-aware and favors reuse.

\section{Evaluation}
\label{sec:eval}
\begin{figure*}[t]
\centering
\includegraphics[width=\textwidth]{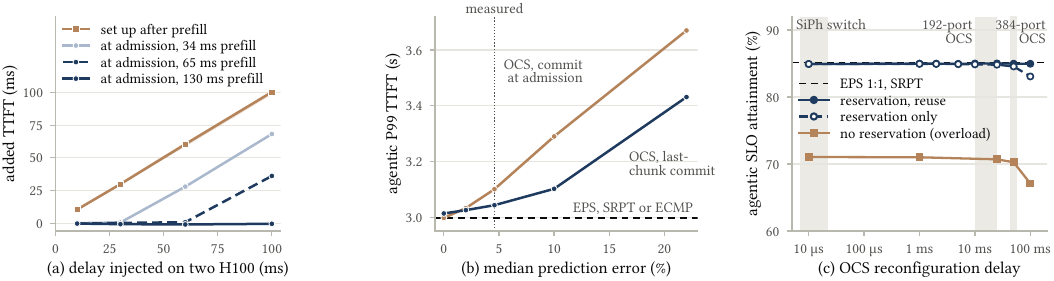}
\caption{(a) TTFT added by a circuit delay injected after prefill or at admission on two H100 GPUs. (b) Agentic
P99 TTFT against median prediction error at 1\,ms and (c) agentic SLO attainment against reconfiguration delay,
simulated at the 128-GPU base point, layer-wise, where no network limit reaches 85\%.}
\label{fig:req}
\end{figure*}

\subsection{Setup}
\emph{Design space.}
A request's KV demand per GPU $D$ is the KV bytes one GPU sends over the prefill time, and $S$ is the effective KV
bandwidth per GPU, so layer-wise streaming exposes $\max(0,1-S/D)$ of TTFT.
Prefill times come from a cost model fitted to 33 vLLM prefill runs of Llama-3.1-70B that we measured on four H100
GPUs (median error 4.6\%), without scaling.
We use H100 GPUs because they are the most advanced GPU system available to us, and the method applies in principle
to newer GPUs by refitting the cost model.

\begin{table}[t]
\centering
\caption{Optical KV cache network requirements and link budget}
\label{tab:req}
\scriptsize
\setlength{\tabcolsep}{2pt}
\newcommand{\two}[1]{\multicolumn{2}{>{\raggedright\arraybackslash}p{0.62\columnwidth}}{#1}}
\begin{tabular}{@{}>{\raggedright\arraybackslash}p{0.34\columnwidth}|>{\raggedright\arraybackslash}p{0.29\columnwidth}|>{\raggedright\arraybackslash}p{0.32\columnwidth}@{}}
\hline
\textbf{Requirement} & \multicolumn{2}{l}{\textbf{Value}} \\
\hline
\textbf{KV bandwidth} per GPU & \two{33\,GB/s for 10\% of mean TTFT} \\
\textbf{Reconfiguration delay} & \two{50\,ms, every tenant within 5\% at 60\% load} \\
\textbf{OCS unit} & \two{384 ports, 50\,ms, 8 planes reach 3,072 GPUs} \\
\textbf{GPU-side optics} & \two{WDM, one fiber pair per 400G link at 45\,GB/s} \\
\hline
\textbf{Optics} & \textbf{DR4} (parallel) & \textbf{WDM} (FR4) \\
\hline
\textbf{OCS ports} per link & 4 or 8 & 1 or 2 \\
\textbf{Cost} over 2:1 EPS & 0.77 to 1.36 & 0.55 to 0.78 \\
\textbf{Break-even port price} & \$305 to \$842 & \$1,050 to \$3,058 \\
\hline
\textbf{Link budget} & \textbf{DR4} & \textbf{WDM} \\
\multicolumn{3}{@{}l}{\emph{one 400G link over 500\,m of fiber, transceiver to OCS to transceiver}} \\
\hline
\textbf{Launch OMA} (dBm) & $-0.8$ to $+4.2$ & $-0.3$ to $+3.7$ \\
\textbf{Rx sensitivity} (dBm) & max($-3.9$, SECQ$-5.3$) & max($-4.6$, SECQ$-6.0$) \\
\textbf{Mux and demux} (dB) & none & 2.5 typ, 3.3 max each \\
\textbf{OCS} (dB) & \two{1.5 typ, 2.7 max} \\
\textbf{Four connectors} (dB) & \two{2.0 typ, 3.0 max} \\
\textbf{Fiber} at 0.5\,dB/km (dB) & \two{0.25} \\
\textbf{Path loss} (dB) & \two{3.75 typ, 5.95 worst} \\
\textbf{Loss allocation} (dB) & 3 & 4 \\
\textbf{Margin} at TDECQ 3.4\,dB & 2.35 typ, 0.15\,dB worst & 2.55 typ, 0.35\,dB worst \\
\textbf{TDECQ} for 2\,dB margin & $\le$1.5\,dB & $\le$1.7\,dB \\
\textbf{TDECQ} for 2\,dB, spliced & $\le$2.4\,dB & $\le$2.6\,dB \\
\hline
\multicolumn{3}{@{}>{\raggedright\arraybackslash}p{0.97\columnwidth}@{}}{\scriptsize Sources are the exposure model, Fig.~\ref{fig:req}(c), the cost model~\cite{mixnet,opus}, data sheets~\cite{polatis7000n,calient,fsdr4,sumitomowd} and specifications~\cite{oifcpo,ieee8023,fr4msa,tia568}. Margins use the maximum launch and the 1550\,nm OCS loss, with mux and demux in the module.} \\
\end{tabular}
\end{table}
\emph{Simulation.}
A flow-level event-driven simulator, driven by request traces as in prior P/D studies~\cite{distserve,splitwise,nuska},
models 128 and 256 H100 GPUs as slices of a larger fabric, with 32-GPU racks and prefill and decode in separate racks.
Each decode instance is a KV sink without decode compute, so the results cover the KV fabric and TTFT.
A TraceLab tenant~\cite{tracelab}, whose decode instances keep no session KV between turns as in Mooncake and
DeepSeek, runs beside Azure 2024 chat and code~\cite{azure24,dynamollm} and Mooncake~\cite{mooncake} tenants, all on
the measured Llama-3.1-70B configuration on four H100 GPUs.
The base point gives TraceLab 25\% of the GPUs at 60\% prefill load, and the grid adds a 50\% share and 75\% load, in
layer-wise and pull transfer.
Results average five paired 330\,s trace windows, with prediction errors drawn from the measured runs.
SLO attainment is the share of requests with TTFT within five times the tenant's median unloaded prefill time.
Without the OCS, the KV crosses a two-tier EPS at 400G or 800G per GPU and at 1:1, 2:1 or 3:1 (3:1 as in
MixNet~\cite{mixnet}, simulated as 2.91:1 with 11 uplinks per 32 GPUs).
The EPS uses ECMP, adaptive routing with fair sharing as in RDMA fabrics~\cite{conga,dcqcn}, or adaptive routing with
global SRPT as a strong packet reference~\cite{pfabric,homa}, and a network without limits gives the ceiling.
The OCS runs last-chunk commit at a 25\,ms delay, and 50\,ms changes its SLO attainment by at most 0.15 percentage
points at 60\% load.
A baseline without reservation holds a circuit as soon as any KV is ready, with the decode chosen by load.

\emph{Cost.}
Both designs keep the same host NIC network, so for new clusters that serve each model within one node, as here, we
compare only the network behind each GPU's 400G RDMA NIC.
Switch ports, NICs and 400G modules use published prices~\cite{mixnet,opus}, and an FR4 module costs 1.47 times a 400G
module as in reseller lists.
The EPS uses the fewest tiers of 51.2T switches, and the OCS one or two ports per fiber pair~\cite{calient,opus} and a
100G control network.

\subsection{Performance and cost}
Migrated turns of Llama-3.1-70B demand a median 28.6\,GB/s per H100 GPU, which fits in a 400G link at 45\,GB/s.
A non-blocking fabric therefore exposes 5\% of mean TTFT, or 10\% at the 32.5\,GB/s NCCL rate we measured over 400G
InfiniBand, and half or a third of 400G per GPU exposes 19\% and 31\% (Fig.~\ref{fig:main}(a)).
At these rates AgentX turns expose 12\% and 20\%, and FP8 KV cuts the TraceLab shares to 5\% and 11\%.

Under load at 128 GPUs, the OCS stays within 0.4 points of the agentic SLO attainment of EPS 1:1 with SRPT at every
grid point and mode, with P99 TTFT within 1.8\%.
Adaptive routing with fair sharing stays within 0.5 points of the OCS even at 3:1, so the loss behind oversubscription
comes from ECMP collisions of migrated turns.
In pull transfer the OCS gains 3.4 to 6.6 points over 2:1 ECMP and 4.6 to 10.7 points over 3:1 ECMP.
In layer-wise transfer it gains 0.2 to 3.2 points and hides 71 to 74\% of migrated turns within 10\% of their prefill
time, against 50 to 58\% for 2:1 and 43 to 53\% for 3:1 ECMP.
At 256 GPUs the pull gains over ECMP are 2.7 to 6.2 points (Fig.~\ref{fig:main}(b)).
The gain follows the share of migrated turns.
When half the turns migrate, as half the decode-side KV does in llm-d~\cite{llmdglm}, the pull gains over 2:1 and 3:1
ECMP are 1.3 to 1.8 and 1.8 to 2.7 points, and under session affinity (7.6\%) every fabric is within 0.6 points, where the OCS keeps its
cost advantage.

At two sets of published prices~\cite{mixnet,opus}, with a NIC and a WDM pluggable per link, the OCS costs 0.55 to
0.78 times a 2:1 EPS and 0.61 to 0.85 times a 3:1 EPS from 256 to 3,072 GPUs, the reach of one OCS stage
(Fig.~\ref{fig:main}(c) shows one price set).
With parallel-fiber DR4 it costs 0.77 to 1.36 times a 2:1 EPS in one stage (Table~\ref{tab:req}), so WDM optics pay
off best.

\subsection{What the optical layer must provide}
\emph{Overlap.}
On our two-H100 testbed (Llama-3.1-8B over NVLink, Fig.~\ref{fig:req}(a)), an injected delay started at admission
stays hidden while it is shorter than the prefill, and one started after prefill adds 0.95 to 1.12 times its length.

\emph{Prediction.}
With last-chunk commit, agentic P99 TTFT rises 1.5\% at the measured 4.6\% median error and 3.6\% at 10\%, against
3.5\% and 9.6\% with admission commit (Fig.~\ref{fig:req}(b)), so P99 stays within 5\% up to 11.6\% error, against
5.9\% with admission commit.

\emph{Reconfiguration delay.}
Without reservation, layer-wise transfers hold decode ports through every prefill.
With the decode chosen by load, the Mooncake tenant then overloads in every trace window even with a 10\,$\mu$s switch
(agentic SLO attainment 71\%), while reservation with the same decode choice holds 85\% at 25\,ms.
Circuit-aware decode choice keeps the same decode for 92\% of consecutive agentic requests, with the busiest decode at
1.07 times the mean, so with reuse the SLO attainment stays at 85\% up to 100\,ms (Fig.~\ref{fig:req}(c)).
Without reuse, all tenants keep P99 and SLO within 5\% of EPS 1:1 with SRPT up to 50\,ms at 60\% load, so a
mechanical OCS suffices.

\emph{Link budget.}
The worst path through four connectors and a 384-port OCS at 2.7\,dB~\cite{polatis7000n} loses 5.95\,dB, above the
3 and 4\,dB channel allocations of DR4 and FR4, so 2\,dB of margin needs a launch near the maximum and a TDECQ of at most 1.5\,dB with DR4 and 1.7\,dB with WDM, or
spliced OCS fibers with typical loss for any compliant TDECQ at the maximum launch.

\section{Related Work}
Optical datacenter networks reconfigure for aggregate traffic~\cite{helios,cthrough} or training
jobs~\cite{sipml,topoopt}, or rotate through traffic-oblivious schedules~\cite{rotornet,sirius}.
MoSE reallocates OCS ports per workload mode without modeling reconfiguration delay~\cite{mose}, xSwitch pairs a
port-reduced EPS with an OCS for scale-up~\cite{xswitch}, a photonic CXL appliance replaces the P/D hop with a memory
pool~\cite{photoniccxl}, and Opus names KV cache pipelines as a target it does not evaluate~\cite{opus}.
KV compression~\cite{cachegen,kivi} loses to raw transfer above 50 to 110\,Gb/s~\cite{kvserve}.

\section{Conclusion}
Committing each circuit during prefill hides the reconfiguration delay of a mechanical OCS, so an optical KV cache
network matches 400G packet fabrics with adaptive routing at 0.55 to 0.78 times the cost of a 2:1 fabric.
In pull transfer it raises agentic SLO attainment over 2:1 and 3:1 ECMP fabrics by 2.7 to 6.6 and 4.1 to 10.7 points.

\end{document}